# Observation of Dirac-like energy band and ring-torus Fermi surface associated with the nodal line in topological insulator CaAgAs

Daichi Takane[1], Kosuke Nakayama[1], Seigo Souma[2,3], Taichi Wada[4], Yoshihiko Okamoto[4,5], Koshi Takenaka[4], Youichi Yamakawa[5,6], Ai Yamakage[5,6], Taichi Mitsuhashi[1,7], Koji Horiba[7], Hiroshi Kumigashira[1,7], Takashi Takahashi[1,2,3], and Takafumi Sato[1,2*]

[1]*Department of Physics, Tohoku University, Sendai 980-8578, Japan*

[2]*Center for Spintronics Research Network, Tohoku University, Sendai 980-8577, Japan*

[3]*WPI Research Center, Advanced Institute for Materials Research, Tohoku University, Sendai 980-8577, Japan*

[4]*Department of Applied Physics, Nagoya University, Nagoya 464-8603, Japan*

[5]*Institute for Advanced Research, Nagoya University, Nagoya 464-8601, Japan*

[6]*Department of Physics, Nagoya University, Nagoya 464-8602, Japan*

[7]*Institute of Materials Structure Science, High Energy Accelerator Research Organization (KEK), Tsukuba, Ibaraki 305-0801, Japan*

## Abstract

One of key challenges in current material research is to search for new topological materials with inverted bulk-band structure. In topological insulators, the band inversion caused by strong spin-orbit coupling leads to opening of a band gap in the entire Brillouin zone, whereas an additional crystal symmetry such as point-group and nonsymmorphic symmetries sometimes prohibits the gap opening at/on specific points or line in momentum space, giving rise to topological semimetals. Despite many theoretical predictions of topological insulators/semimetals associated with such crystal symmetries, the experimental realization is still relatively scarce. Here, using angle-resolved photoemission spectroscopy with bulk-sensitive soft x-ray photons, we experimentally demonstrate that hexagonal pnictide CaAgAs belongs to a new family of topological insulators characterized by the inverted band structure and the mirror reflection symmetry of crystal. We have established the bulk valence-band structure in three-dimensional Brillouin zone, and observed the Dirac-like energy band and ring-torus Fermi surface associated with the line node, where bulk valence and conducting bands cross on a line in the momentum space under negligible spin-orbit coupling. Intriguingly, we found that no other bands cross the Fermi level and therefore the low-energy excitations are solely characterized by the Dirac-like band. CaAgAs provides an excellent platform to study the interplay among low-energy electron dynamics, crystal symmetry, and exotic topological properties.

## Introduction

Topological insulators (TIs) exhibit a novel quantum state with metallic edge or surface state (SS) within the bulk band gap generated by the strong spin-orbit coupling (SOC). The topological SS in three-dimensional (3D) TIs is characterized by a linearly dispersing Dirac-cone energy band,[1-3] which hosts massless Dirac fermions protected by the time-reveal symmetry (TRS). The discovery of TIs triggered the search for new types of topological materials containing surface or bulk Dirac-cone bands protected by crystal symmetries, as represented by topological crystalline insulators (TCIs) with the Dirac-cone SSs protected by mirror symmetry,[4-6] as well as 3D Dirac semimetals (DSMs) with bulk Dirac-cone bands protected by rotational symmetry (such as $Cd_3As_2$ and $Na_3Bi$).[7-11] While the Dirac cone in DSMs is spin degenerate, breaking the TRS or space-inversion symmetry (SIS) leads to the Weyl-semimetal (WSM) phase with pairs of spin-split Dirac (Weyl) cones, as recently verified in transition-metal monopnictides.[12-14] Such Dirac-cone states are known to provide a platform to realize outstanding physical properties such as extremely high mobility, gigantic linear magnetoresistance, and chiral anomaly.[15-22]

While the DSMs and WSMs are characterized by the crossing of bulk bands at the discrete points in $k$ space (point nodes), there exists another type of topological semimetal characterized by the band crossing along a one-dimensional curve in $k$ space (line node), called line-node semimetal (LNSM). The LNSMs are expected to show unique physical properties different from the DSMs and WSMs, such as a flat Landau level, the Kondo effect, long-range Coulomb interaction, and peculiar charge polarization and orbital magnetism.[23-26] Despite many theoretical predictions of LNSMs in various material platforms,[27-54] experimental studies on the LNSMs are relatively scarce.[55-62]

Recently, it was theoretically proposed by Yamakage *et al*. that noncentrosymmetric ternary pnictides CaAgX (X = P, As) are the candidate of LNSM and TI.[41] These materials crystalize in the ZrNiAl-type structure with space group $P\bar{6}2m$ (No. 189)[63] (for crystal structure, see Fig. 1a). First-principles band-structure calculations have shown that, under negligible spin-orbit coupling (SOC), CaAgX displays a fairly simple band structure near the Fermi level ($E_F$) with a ring-like line node (nodal ring) surrounding the Γ point of bulk hexagonal Brillouin zone (BZ) (bulk BZ is shown in Fig. 1b). The line node is associated with the crossing of bulk conduction band (CB) and valence band (VB) with Ag $s$ and P/As $p$ character, respectively, and is protected by the mirror reflection symmetry of crystal. When the SOC is included in the calculation, CaAgP still keeps the line node due to the very small spin-orbit gap (~ 1 meV) while a relatively large spin-orbit gap (~ 75 meV) opens along the line node in CaAgAs to make this material a narrow-gap TI.[41] The SOC thus plays a crucial role in switching the LNSM and TI phases in CaAgX. Transport measurements on the CaAgP and CaAgAs polycrystalline samples by Okamoto *et al*. have demonstrated the low-carrier-density nature of these samples, consistent with the existence of line node[64] (note that this is further corroborated by the recent transport measurements on CaAgX single crystals[62]). They have further suggested that $E_F$ in hole-doped CaAgAs sample lies in the middle of linearly dispersive Dirac-like band, and concluded that CaAgAs is well suited to study the low-energy excitations related to the Dirac electrons. While the electronic states of CaAgX and its relationship to physical properties should be experimentally examined, there exist no experimental outputs on the band structure of CaAgX. It is thus urgently required to establish the fundamental electronic states.

In this work, we report the ARPES study of CaAgAs. By utilizing bulk-sensitive soft-x-ray photons from synchrotron radiation, we established the bulk VB structure in the 3D bulk BZ. We suggest that CaAgAs is a narrow-gap TI with an ideal band structure suitable to study the low-energy excitations linked to the bulk Dirac-like band arising from the line node. This is demonstrated by observing the bulk Fermi surface which is *solely* derived from the VB and CB associated with the *single* line node, consistent with our first-principles band structure calculations. We discuss the consequence of our observation in relation to the exotic physical properties.

**Results and Discussion**

Samples and experimental

High-quality single crystals of CaAgAs were grown on the sintered pellets of CaAgAs (for details, see Method). A typical photograph of our single crystal is shown in Fig. 1c. ARPES measurements were performed with synchrotron light at BL2 in Photon Factory, KEK. Samples were cleaved *in situ* along the ($11\bar{2}0$) crystal plane (a shiny mirror plane in Fig. 1c) as confirmed by the Laue x-ray diffraction measurement on the cleaved surface (typical Laue pattern is shown in Fig. 1d) and the photon-energy ($h\nu$) dependence of the band dispersion. This indicates that the cleaved plane is the $k_y$ - $k_z$ plane in the hexagonal BZ (Fig. 1b). Figure 1e displays the energy distribution curve (EDC) in the wide energy region measured at $h\nu$ = 580 eV. One can recognize several core-level peaks originating from the Ca (3*s*, 3*p*), Ag (4*s*, 4*p*, 4*d*), and As (3*s*, 3*p*, 3*d*) orbitals. No other core-level peaks were found in this energy range, confirming the clean sample surface.

Valence-band structure

First, we present the overall VB structure of CaAgAs. We found that soft-x-ray photons are useful for revealing the bulk electronic states of CaAgAs as in the case of noncentrosymmetric Weyl semimetals such as TaAs,[12-14] although we need to sacrifice the energy/momentum resolution compared to the vacuum ultraviolet (VUV) photons. In fact, the obtained VUV data were found to suffer large broadening along wave vector perpendicular to the surface probably because of the final-state effect and rather rough nature of the cleaved surfaces, and therefore we concluded that the VUV photons are not best suited for resolving 3D electronic states of CaAgAs. Figure 2a displays the EDCs at the normal emission measured with various photon energies in the soft-x-ray region of 530 – 700 eV. One can identify several dispersive bands. For example, a band located at binding energy ($E_B$) of ~ 1 eV at $h\nu$ ~ 700 eV, which is attributed to the topmost bulk VB, disperses toward $E_F$ on decreasing $h\nu$ to 650 eV, and it disperses back again toward ~ 1 eV at $h\nu$ = 600 eV. On further decreasing $h\nu$ down to ~ 550 eV, this band again approaches $E_F$, similarly to the case of $h\nu$ = 650 - 700 eV. Such periodic band dispersion was also observed for the bands located at $E_B$ ~1.5 and ~3 eV. To visualize the experimental band dispersion more clearly, we plot in Fig. 2b the band structure obtained from the second derivative of the ARPES intensity plotted as a function of wave vector perpendicular to the sample surface ($k_x$), which corresponds to the ΓM cut in the bulk BZ (cut A in Fig. 2c). One can immediately recognize that the overall experimental band dispersion shows a reasonable agreement with the calculated bulk bands (red curves) regarding the periodicity and location of bands; this confirms that the cleaving plane is (11$\bar{2}$0). The holelike dispersion approaching $E_F$ around the Γ point is well reproduced by the calculations, and therefore it is assigned as the topmost VB with the As 4$p$ orbital character. Moreover, a good agreement of the band width between experiment and calculation signifies no apparent band-renormalization effect, suggesting the weak electron correlation. It is noted that we observe a single holelike band within 1.5 eV of $E_F$ in the experiment, while the calculation predicts two holelike bands. Such difference may be due to the finite *k*/energy-broadening effect as well as the matrix-element effect of photoelectron intensity which turned out to be rather strong in this material.

We comment here that our Hall conductivity measurement of CaAgAs single crystal suggests the existence of hole carriers with carrier concentration of ~ 1.6 × $10^{20}$ cm$^{-3}$, which corresponds to the $E_F$ location of 0.27 eV below the VB top. On the other hand, the VB structure determined by ARPES shows a reasonable agreement with the calculated band structure without sizable $E_F$ shift. While we do not know the exact origin of such a difference, some possibilities like downward surface band-bending may be considered.

To see the band dispersion along the in-plane wave vector, we used the photon energy of $h\nu$ = 550 eV which traces the *k* cut crossing the Γ point of 13th BZ, and measured the EDC along the ΓKM ($k_y$) cut (cut B in Fig. 2c), as shown in Fig. 2d. One can recognize highly dispersive holelike band centered at the Γ point, similarly to the case of ΓM cut in Figs. 2a and 2b. This band is better visualized in the ARPES-intensity plot in Fig. 2e in which a holelike band with linear dispersion shows up in the $E_B$ range of 0 ~ 1.5 eV (note that the intensity distribution around the M and Γ points is different between Figs. 2b and 2e and also among different BZs due to aforementioned matrix-element effect). We have confirmed by the band-structure mapping in 3D BZ that the Fermi surface exists only around the Γ point. This conclusion is supported by the experimental band dispersion along the AHL cut in Fig. 2f (cut C in Fig. 2c) where the topmost VB always stays below $E_B \sim$ 1 eV without crossing $E_F$.

Ring-torus Fermi surface

Having established the overall VB structure, a next important issue is the electronic structure in the vicinity of $E_F$ responsible for the physical properties. Figure 3a displays the ARPES intensity at $E_F$ as a function of $k_x$ and $k_y$ (the ΓKM plane). One immediately finds a bright intensity pattern surrounding the Γ point, in particular in 13th BZ, confirming the absence of additional Fermi surface away from the Γ point. It is also obvious from Fig. 3b that no Fermi surface exists away from Γ in the $k_y$-$k_z$ (ΓAKH) plane. As shown in Figs. 3a and 3b, when we overlaid the calculated Fermi surface (green curves) onto the ARPES intensity (note that we assumed the location of $E_F$ to be 0.05 eV below the VB top in the calculation to account for a small but finite hole-doping effect in experiment), the high-intensity region coincides with the $k$ region where the calculated Fermi surface exists.

To gain further insight into the Fermi-surface topology, we show in the top panels of Figs. 3c and 3d the ARPES intensity near $E_F$ and the intensity obtained by taking second derivative of the EDCs, respectively, along the $k$ cut nearly crossing the Γ point (cut A in Fig. 3a). One finds a linearly dispersive holelike band originating from the As 4$p$ states which is better visualized in the second-derivative plot in Fig. 3d (by a linear extrapolation of the band dispersion around $E_F$, we have estimated the Fermi velocity to be $v_F = 2.1 \pm 0.1$ eVÅ). This band is reproduced by our calculation as shown by red curves in Fig. 3c, and is responsible for the outer ring in Fig. 3a. As shown in cut A of Fig. 3c, there exists another electronlike band in the calculation which originates from the Ag 5$s$ orbital; this band forms the inner ring in Fig. 3a. While the intensity of the electronlike band seems weak in the original intensity (Fig. 3c), the second-derivative image in Fig. 3d shows a finite spectral weight likely arising from the electronlike band.

As shown in cut A of Fig. 3c, the calculated electronlike band intersects the holelike band at ~ 0.1 eV above $E_F$, and forms the nodes at $k_y$ ~ ± 0.15 Å$^{-1}$ under negligible SOC, since cut A is on the (0001) mirror plane (the $k_x$-$k_y$ plane) and the nodes are protected by mirror reflection symmetry.[41] This indicates that the electronic states within two opposite nodes across the Γ point have an inverted band character. Thus, our observation of electronlike feature can be regarded as a hallmark of the band inversion, which is a prerequisite for realizing LNSM or TI. It is remarked that with a finite SOC, an energy gap of 75 meV opens in the caluclation, as can be seen from a difference in the band dispersion with (solid curves) and without (dashed curves) SOC in Fig. 3c.[41] The opening of a spin-orbit gap at the node is also seen in some other LNSM candidates such as $Cu_3(Pd,Zn)N$,[34,35] $Ca_3P_2$,[37,49] ZrSiS,[40] CaTe,[50] and fcc alkaline-earth metal[51].

To clarify whether the node-like feature in CaAgAs is seen at a point or on a line in $k$ space, it is necessary to measure the band dispersion along different $k$ slices around the Fermi surface. For this sake, we show in the middle and bottom panels of Figs. 3c and 3d the intensity for cuts slightly away from the Γ point (cuts B and C in Fig. 3a) obtained with different $hv$'s. One can recognize that overall band structure along cut B is similar to that along cut A regarding the $E_F$ crossing of holelike band and the presence of electronlike feature. This is reasonable since cut B is also on the mirror plane and still crosses the calculated nodal points. On the other hand, along cut C, the holelike band moves downward and shows no $E_F$ crossing. These behaviors are consistent with the presence of a ring-shaped nodal feature (nodal ring) on the mirror plane shown by a dashed curve in Fig. 3a.

It should be stressed again that there exists a spin-orbit gap along the nodal line in the calculation. Since the gap is almost isotropic (75±1 meV) along the nodal ring (not shown), the low-energy excitations in CaAgAs are characterized by the excitations across the band gap in the $k$ region involving the entire line node. Unfortunately, such a band gap (as well as the topological SSs) was not resolved in the ARPES experiment, likely due to the slightly hole-doped nature of crystal. Considering the fact that (i) the calculated spin-orbit gap is not so small compared to other TIs and (ii) the ARPES-derived band dispersion shows a reasonable agreement with the calculation near $E_F$, it would be more reasonable to regard CaAgAs as a narrow-gap TI, rather than a LNSM. It is also emphasized here that the TI nature of CaAgAs should be distinguished from that of prototypical TIs such as $Bi_2Se_3$ since the node never shows up even without SOC in $Bi_2Se_3$ unlike CaAgAs.

Figure 3e illustrates the calculated equi-energy contour maps in $k$ space for selected energy slices (energy with respect to the VB top, $E_{VB}$ = 0.1, 0.27, and 0.6 eV). The energy contour around $E_F$ ($E_{VB}$ = 0.1 eV) shows a ring-torus shape. On increasing $E_{VB}$, it gradually expands and fattens. Eventually it transforms into the spheroid as soon as the $E_{VB}$ passes the bottom of electonlike band. As shown in Fig.

3f, such evolution of the ring-torus contours can be seen from the ARPES-intensity mapping for representative $E_B$ slices compared with the calculated energy contours.

From all these experimental results, we concluded that CaAgAs is likely a narrow-gap TI characterized by the bulk Dirac-like band and ring-torus Fermi surface associated with the line node on the (0001) mirror plane, in line with the first-principles band-structure calculations. Intriguingly, we revealed that the bulk Fermi surface is solely derived from the Dirac-like bands associated with a single line node (which appears under negligible SOC), and no other bands cross $E_F$. In this regard, the CaAgX family can be distinguished from some other LNSM candidates which contain additional normal bands crossing $E_F$ and multiple line nodes[55-61]. Thus, CaAgX is a promising platform to study the interplay among mirror symmetry, low-energy excitations, and transport properties in LNSMs. Also, it would provide an excellent platform to study the influence of SOC to the low-energy excitations involving the line node, because the SOC strength can be controlled by replacement of P and As in the crystal. Since the bulk electronic states of CaAgAs have been established in this study, a next important challenge is to clarify the nature of predicted topological SSs in the band inverted region.[33] Such an experiment would require an access to the band dispersion above the line node *via* the fabrication of electron-rich CaAgAs.

**Methods**

Sample preparation

High-quality single crystals of CaAgAs were synthesized by the following procedure. An equimolar mixture of calcium chips, silver powder, and arsenic chunks were put in an alumina crucible and sealed in an evacuated quartz tube. The tubes were kept at 773 K for 12 h and then at 1273 K for 12 h, followed by furnace cooling to room temperature. The obtained samples were pulverized, pressed into pellets, and sealed in quartz tubes. The pellets were sintered at 1173 K for 2 h and cooled to room temperature at a rate of 30 K h$^{-1}$, resulting in that shiny hexagonal-prismatic single crystals of CaAgAs were grown on the pellets. The quality of the crystal was checked by X-ray diffraction technique using a RIGAKU R-AXIS IP diffractometer.

ARPES experiments

ARPES measurements were performed with a Scienta-Omicron SES2002 electron analyzer with energy-tunable synchrotron light at BL2 (Multiple Undulator beamline for Spectroscopic Analysis on Surface and HeteroInterface; MUSASHI) in Photon Factory, KEK. We used linearly polarized light (horizontal polarization) of 500-975 eV. The energy and angular resolutions were set at 150 meV and 0.2°, respectively. Samples were cleaved *in situ* in an ultrahigh vacuum better than 1×10$^{-10}$ Torr along the $(11\bar{2}0)$ crystal plane, as confirmed by the Laue x-ray diffraction measurement on the cleaved surface and the photon-energy dependence of the band dispersion shown in Fig. 2b. Sample temperature was kept at $T = 40$ K during the ARPES measurements. The Fermi level ($E_F$) of samples was referenced to that of a gold film evaporated onto the sample holder.

Calculations

Electronic band-structure calculations were carried out by means of first-principles band structure calculations by using WIEN2k code[65] with the full-potential linearized augmented plane-wave method within the generalized gradient approximation. We used the experimental structural parameters for the calculations.[63] 24 × 24 × 36 $k$-points sampling was used for the self-consistent calculations.[41]

Data availability

The data that support the findings of this study are available from the corresponding author upon reasonable request.

**Acknowledgements**

We thank H. Oinuma and T. Nakamura for their assistance in the ARPES measurements and T. Yajima for his assistance in the single-crystal X-ray diffraction experiments. This work was supported by Grant-in-Aid for Scientific Research on Innovative Areas "Topological Materials Science" (JSPS KAKENHI No: JP15H05853), Grant-in-Aid for Scientific Research (JSPS KAKENHI No: JP17H01139, JP15H02105, JP26287071, JP25287079, JP25220708, JP16K13664, and JP16K17725), KEK-PF (Proposal No: 2015S2-003 and 2016G555), and UVSOR (Proposal No: 28-542 and 28-828).

## Author Contributions

The work was planned and proceeded by discussion among D.T., A.Y., Y.O. and T.S., and T.W., Y.O., and K.T. carried out the samples' growth and their characterization. D.T., K.N. S.S. T.M., K.H., H.K, T.T., and T.S. performed ARPES measurements. Y.Y. and A.Y. performed the first-principles band structure calculations. D.T. and T.S. finalized the manuscript with inputs from all the authors.

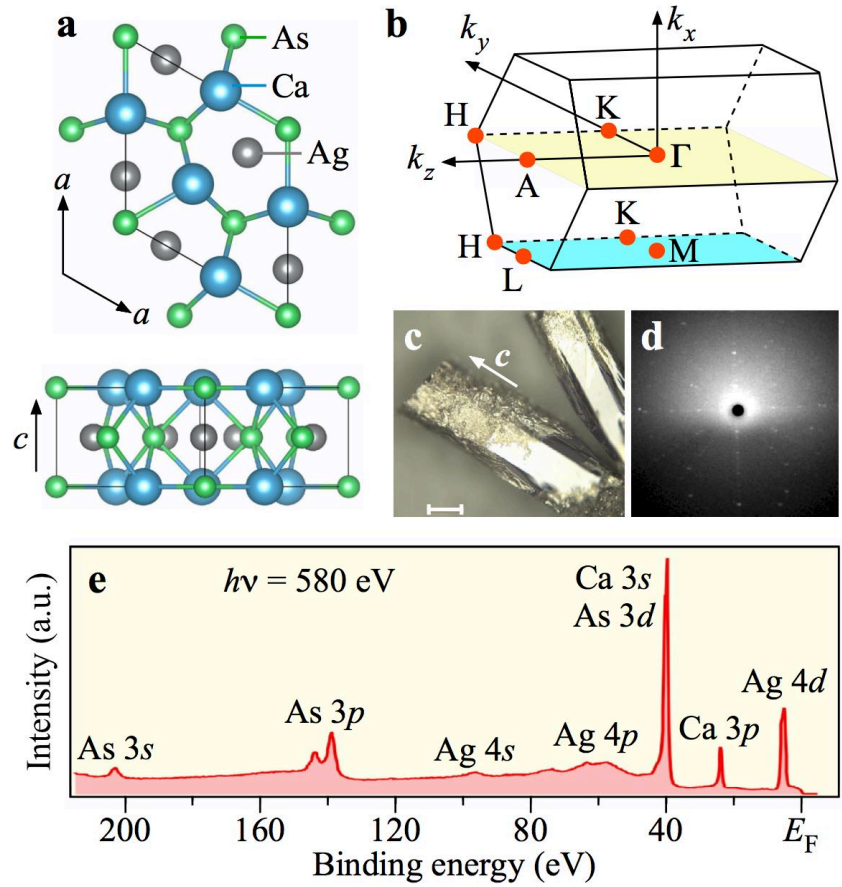

Fig. 1 Characterization of CaAgAs single crystal. **a** Two different views of crystal structure of CaAgAs. **b** Bulk Brillouin zone of CaAgAs. **c** Photograph of CaAgAs single crystal. Scale bar: 200 μm. **d** Laue x-ray diffraction pattern for the cleaved surface. **e** The EDC in the wide energy region measured at $T = 40$ K with $h\nu = 580$ eV.

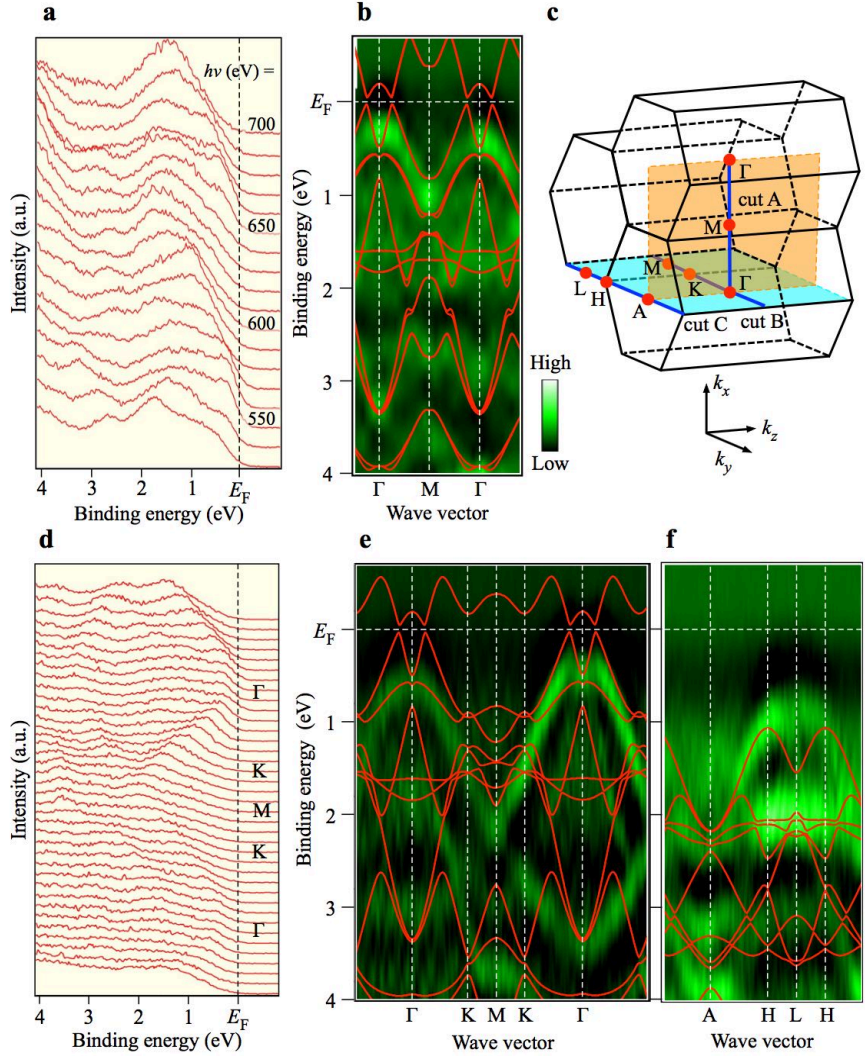

Fig. 2 Valence-band structure of CaAgAs. **a** Photon-energy dependence of normal-emission EDCs in the VB region. **b** ARPES intensity in the VB region along the ΓM line (cut A) plotted as a function of wave vector ($k_x$) and $E_B$, together with the calculated band dispersions (red curves).[41] ARPES intensity was obtained by taking second derivative of the EDCs. **c** Bulk BZ and measured $k$ cuts. **d** The EDCs along the ΓKM line (cut B), and **e** corresponding second-derivative ARPES intensity. **f** Same as **e** but measured along the AHL line (cut C).

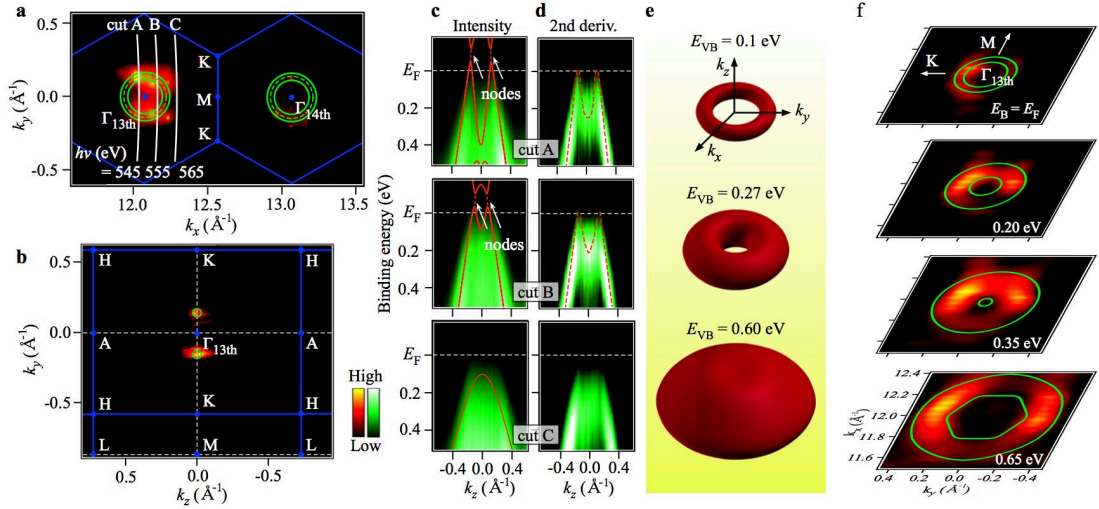

Fig. 3 Line-node semimetallic nature of CaAgAs. **a** ARPES-intensity mapping at $E_F$ as a function of $k_x$ and $k_y$, together with the calculated Fermi surface (solid green curves). We set the $E_F$ position in the calculation to be 0.05 eV below the VB top, to take into account slight hole-doping effect. Calculated ring-like line node (nodal ring) is shown by a dashed circle in BZ. **b** Same as **a** but as a function of $k_y$ and $k_z$ (ΓAHK plane). **c** and **d** ARPES intensity and the intensity obtained by taking second derivative of the EDCs, respectively, measured along three representative cuts (cuts A-C in **a**). Calculated bands with SOC are shown by solid red curves in **c**, while those without SOC are indicated by dashed curves around the node. Dashed red curves in **d** are a guide to the eyes to trace the experimental band dispersions. **e** Calculated equi-energy contour maps in $k$ space for selected energy slices (energy with respect to the VB top, $E_{VB}$ = 0.1, 0.27, and 0.6 eV). **f** ARPES-intensity mapping as a function of two-dimensional wave vector ($k_x$ and $k_y$) for representative $E_B$ slices compared with the calculated energy contours (green curves). ARPES intensity in **a**, **b**, and **f** was obtained by taking second derivative of the EDCs.